\documentstyle[12pt,aaspp4]{article}

\begin{document}

\title{DETECTION OF MULTI-TeV GAMMA RAYS FROM MARKARIAN 501 DURING  AN UNFORESEEN 
FLARING STATE IN 1997 
WITH THE TIBET AIR SHOWER ARRAY}

\author{ M.~Amenomori\altaffilmark{1}, S.~Ayabe\altaffilmark{2}, 
        P.Y.~Cao\altaffilmark{3},       Danzengluobu\altaffilmark{4},
        L.K.~Ding\altaffilmark{5},     Z.Y.~Feng\altaffilmark{6}, 
        Y.~Fu\altaffilmark{3},          H.W.~Guo\altaffilmark{4}, 
        M.~He\altaffilmark{3},          K.~Hibino\altaffilmark{7},
        N.~Hotta\altaffilmark{8},       Q.~Huang\altaffilmark{6},
        A.X.~Huo\altaffilmark{5},       K.~Izu\altaffilmark{9},
        H.Y.~Jia\altaffilmark{6},       F.~Kajino\altaffilmark{10}, 
        K.~Kasahara\altaffilmark{11},   Y.~Katayose\altaffilmark{9},
        Labaciren\altaffilmark{4},      J.Y.~Li\altaffilmark{3},
        H.~Lu\altaffilmark{5},          S.L.~Lu\altaffilmark{5},
        G.X.~Luo\altaffilmark{5},       X.R.~Meng\altaffilmark{4}, 
        K.~Mizutani\altaffilmark{2},     J.~Mu\altaffilmark{12}, 
        H.~Nanjo\altaffilmark{1},       M.~Nishizawa\altaffilmark{13},
        M.~Ohnishi\altaffilmark{9},     I.~Ohta\altaffilmark{8},
        T.~Ouchi\altaffilmark{7},       J.R.~Ren\altaffilmark{5},
        T.~Saito\altaffilmark{14},      M.~Sakata\altaffilmark{10},
        T.~Sasaki\altaffilmark{10},     Z.Z.~Shi\altaffilmark{5},
        M.~Shibata\altaffilmark{15},    A.~Shiomi\altaffilmark{9}, 
        T.~Shirai\altaffilmark{7},      H.~Sugimoto\altaffilmark{16},
        K.~Taira\altaffilmark{16},      Y.H.~Tan\altaffilmark{5},
        N.~Tateyama\altaffilmark{7},    S.~Torii\altaffilmark{7}, 
        T.~Utsugi\altaffilmark{2},      C.R.~Wang\altaffilmark{3},
        H.~Wang\altaffilmark{5},        
        X.W.~Xu\altaffilmark{5},        Y.~Yamamoto\altaffilmark{10},
        G.C.~Yu\altaffilmark{6},        A.F.~Yuan\altaffilmark{4}, 
        T.~Yuda\altaffilmark{9},        C.S.~Zhang\altaffilmark{5},
        H.M.~Zhang\altaffilmark{5},     J.L.~Zhang\altaffilmark{5},
        N.J.~Zhang\altaffilmark{3},     X.Y.~Zhang\altaffilmark{3}, 
        Zhaxiciren\altaffilmark{4},     Zhaxisangzhu\altaffilmark{4},
        and  W.D.~Zhou\altaffilmark{12}
         (The Tibet AS${\bf \gamma}$ Collaboration)}

\altaffiltext{1}{ Department of Physics, Hirosaki University, Hirosaki 036-8561, Japan}
\altaffiltext{2}{ Department of Physics, Saitama University, Urawa 338-8570, Japan}
\altaffiltext{3}{ Department of Physics, Shangdong University, Jinan 250100, China}
\altaffiltext{4}{  Department of Mathematics and Physics, Tibet University, Lhasa 850000, China}
\altaffiltext{5}{ Institute of High Energy Physics, Academia Sinica, Beijing 100039, China}
\altaffiltext{6}{  Department of Physics, South West Jiaotong University, Chengdu 610031, China}
\altaffiltext{7}{  Faculty of Engineering, Kanagawa University, Yokohama 221-8686, Japan}
\altaffiltext{8}{  Faculty of Education, Utsunomiya University, Utsunomiya 321-8505, Japan}
\altaffiltext{9}{  Institute for Cosmic Ray Research, University of Tokyo, Tanashi 188-8502, Japan}
\altaffiltext{10}{ Department of Physics, Konan University, Kobe 658-8501, Japan}
\altaffiltext{11}{ Faculty of Systems Engineering, Shibaura Institute of Technology, Omiya 330-8570, Japan}
\altaffiltext{12}{ Department of Physics, Yunnan University, Kunming 650091, China}
\altaffiltext{13}{ National Center for Science Information Systems, Tokyo 112-8640, Japan}
\altaffiltext{14}{ Tokyo Metropolitan College of Aeronautical Engineering, Tokyo 116-0003, Japan}
\altaffiltext{15}{ Faculty of Engineering, Yokohama National University, Yokohama 240-0067, Japan}
\altaffiltext{16}{ Shonan Institute of Technology, Fujisawa 251-8511, Japan}

\begin{abstract}

 In 1997, the BL Lac Object Mrk 501 entered  a very active phase and was the brightest 
source in the sky at TeV energies, showing strong and frequent flaring.  
Using the data obtained with a high density air shower array that has been 
operating successfully at Yangbajing in Tibet since 1996, we searched for $\gamma$-ray signals 
from this source during the period from February through  August in 1997.  
Our observation detected multi-TeV $\gamma$-ray signals at the 3.7~$\sigma$ level 
during this period. The most rapid increase  of the excess counts was observed 
between April 7 and June 16 and  the statistical significance of the excess counts
 in this period was  4.7~$\sigma$.   Among several observations of flaring TeV 
$\gamma$-rays from Mrk 501 in 1997, this is the only observation  using a conventional 
air shower array. We present the energy spectrum of $\gamma$-rays which will be worthy 
to compare with those obtained by imaging atmospheric Cerenkov telescopes. 

\end{abstract}


\section{INTRODUCTION}

 Mrk 501 and Mrk 421 have been well detected as extra-galactic TeV $\gamma$-ray sources by
Whipple and subsequent ground-based Cerenkov detectors (\cite{ong98}). They are the 
so-called BL Lac objects, which are radio-loud AGNs (Active Galactic Nuclei) whose 
relativistic jets are aligned along our line of sight. Flux variability on various 
scales is a common feature of BL Lac objects as already seen in Mrk 501 and Mrk 421, 
and spectral variations of $\gamma$-rays coming from these sources are considered to be
a very powerful tool for understanding the physics of BL Lac objects.
When Mrk 501 was first detected by the Whipple Collaboration in 1995 (\cite{quin96}), 
it showed rather low fluxes at a level significantly below the Crab flux. In March of 1997, 
however, this source went into a state  of remarkably flaring activity and its high state 
lasted for almost half a year with highly variable and strong $\gamma$-ray emission.  
The maximum flux  reached roughly 10 times that of the Crab. During this period, 
several groups (\cite{prot97}) observed strong $\gamma$-ray emission from this source 
with imaging atmospheric Cerenkov detectors.  Independent measurements of the 
$\gamma$-ray spectrum  seem to show a gradual softening towards higher energy, while  
the systematic uncertainties in the flux estimates remain too large to reach a common 
understanding. The energy spectrum and its shape are very important quantities for clarifying the
mechanism of $\gamma$-ray production or particle acceleration at the source, and
 eventually to lead to the actual measurement  of the intergalactic infrared or 
far-infrared background field (\cite{steck93}). Hence,  confirmation of the detection of $\gamma$-rays 
with a different technique will be strongly required.

The Tibet air shower array, operating since 1990, is located at Yangbajing (4300 m above
sea level) in Tibet (\cite{ame92}). This array has a capability of detecting $\gamma$-rays 
in the TeV energy region with high efficiency and good angular resolution. Using  
this array, we  have succeeded in detecting the Crab at the 5.5~$\sigma$ level (\cite{ame99}). 
In this paper we  present the observation of multi-TeV  $\gamma$-ray flares  from Mrk 501 
in 1997. The result obtained with well established air shower technique is important for 
comparing  with those by imaging atmospheric  Cerenkov telescopes.  

\section{EXPERIMENT}

 The Tibet air shower array consists of two overlapping arrays (Tibet-II and HD) as 
described elsewhere (\cite{yuda96}).  The Tibet-II array comprises 185 scintillation 
detectors (BICRON 408A) of 0.5 m$^2$ each  placed on a 15 m square  grid with an 
enclosed area of 36,900 m$^2$, and the HD (high density) array is operating inside 
the Tibet-II array to detect cosmic ray showers with energies lower than 10 TeV.   This HD array 
consists of  109 scintillation detectors (some of detectors are commonly used in 
both arrays),  placed on a 7.5 m square grid covering an area of 5,175 m$^2$.  
The detector arrangement of the Tibet air shower array is schematically shown in 
Fig. 1.  Every detector, except those placed with a 30 m spacing on the
outskirts of the inner detector matrix of the Tibet-II array, is equipped with a
 fast timing (FT) phototube (HPK H1161) and is thus referred to as an ^^ ^^ FT-detector''.
 A lead plate of 5 mm thickness is placed on the top 
of each detector to improve  the fast timing data by converting $\gamma$-rays in the 
showers to electron pairs. This lead converter typically increases the shower size 
by a factor of about 2 and improves the angular resolution by about 30 \% (\cite{ame90}). 

\begin{figure}
\epsscale{.6}
\plotone{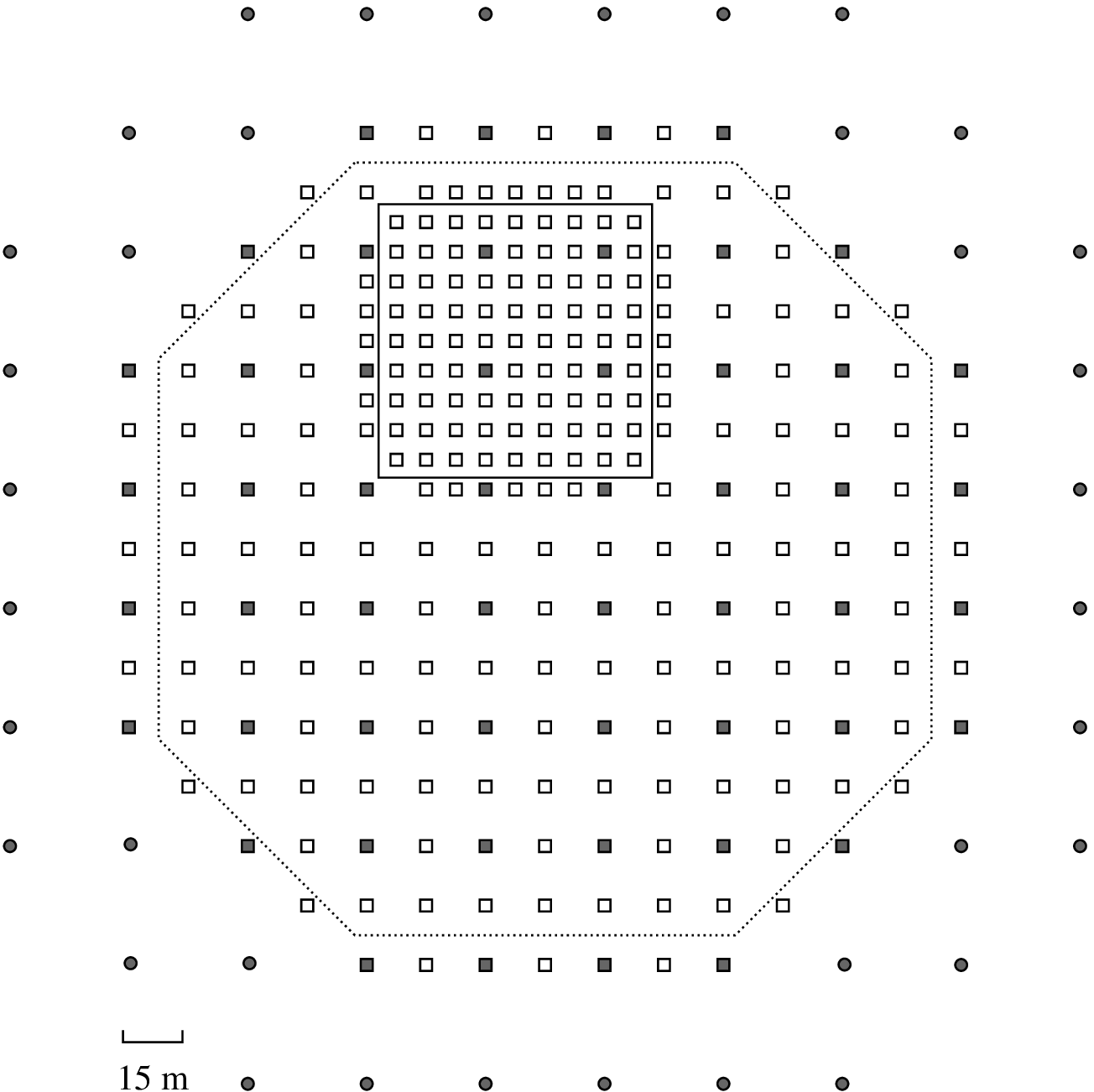}
\caption{Schematical view of the Tibet-II/HD  shower array operating at Yangbajing.  
Open and filled squares : FT-detectors ; filled  circles : density detectors equipped with 
wide dynamic range phototube.  We selected the events whose cores are within the detector 
matrix enclosed with the solid (HD)  or dotted (Tibet-II) line. \label{fig1}}
\end{figure}

\placefigure{fig1}

All the TDCs (time-to-digital converters) and ADCs (analog-to-digital converters) are 
regularly monitored  by using a calibration  module in the FASTBUS system at every 
20 minutes. The length of each signal cable is also monitored by measuring a 
mismatched-reflection pulse from each detector. 
 The data-taking  system has been operating under any 4-fold coincidence in the
FT-detectors, resulting in that the trigger rate of the events is  about 200~Hz for
the Tibet-II while being about 115~Hz for the HD array. 

The observation presented here was made by using the data taken between
  1997 February and 1997 August. The event selection was done by  imposing the 
following three conditions to the recorded  data : 1) Each of any four FT detectors 
should record a signal of more than 1.25 particles ; 2) among the four detectors recording the
highest particles, two or more should be within each detector area of the Tibet II 
and HD arrays denoted  by the dotted and solid lines, respectively, in Fig.1 ;
 and 3) the zenith angle of the incident direction should be less than 45$^\circ$.
After data processing and quality cuts, the total number of events selected  were
$5.5 \times 10^8$  for the HD array and  $1.0  \times 10^9$  for the Tibet-II array,
 respectively, with the effective running time of 155.3 days.

\section{ARRAY PERFORMANCE}

  Since the background cosmic rays are isotropic and $\gamma$-rays from a source are 
apparently centered on the source direction, a bin size for collecting on-source  
data should be determined based on the array's angular resolution so as to optimize 
the signal to noise ratio. In order to achieve a good resolution, a study of 
core-finding techniques and shower-front curvature corrections has been done (\cite{ame90}).
The  angular accuracy of the Tibet array can be  checked thoroughly by observing the shadow 
that the Moon casts in the cosmic rays (\cite{ame93}).
The Tibet II and HD arrays have a capability of observing the Moon's shadow with good statistics. 
 The mode energies  of primary protons to be detected are about 3 TeV and about 8 TeV 
for the Tibet HD and II arrays, respectively. 
Hence the angular resolution of each array can be independently examined in respective 
energy regions. The statistical significance of the Moon's shadow observed with both arrays 
becomes about 10$\sigma$ or more for half a year observation. From this observation,
 we estimated the angular resolution of both  arrays to be better than 0.9$^\circ$ for all events. 
We have also found that the angular resolution scales with $\sum \rho$, where 
 $\sum \rho$ stands for  
the sum of the number of shower particles per {\rm m$^2$} detected in each counter.
    The resolution  increases with increasing $\sum\rho$ as
 $0.8^\circ \times ((\ge\sum\rho)/20)^{-0.3}$ ( $15 < \sum \rho < 300$ ).

  The Moon's shadow by the events with $\sum\rho$ = 15-50 was found at the position 
shifted from the Moon center to the west by 0.32$^\circ$ ($\pm 0.10^\circ$). 
 The primary cosmic rays 
casting the Moon's shadow are almost protons and the mean energy of protons capable of 
generating these events at Yangbajing  is estimated to be about 4.7 TeV by the simulation.
On the other hand, a proton of energy E impinging at normal angle on the Earth is 
deflected by the geomagnetic field and its deflection angle is calculated 
as $ \Delta \theta  E \simeq 1.6^\circ$ TeV. 
So, the observed shift of the Moon's shadow is consistent with that expected from
 the effect of
the geomagnetic field.  A more elaborate study of the Moon's shadow using a Monte 
Carlo technique shows almost same results as those by the experiment
 (\cite{suga99}).  
Thus, the results obtained by assigning primary energies to  the
observed events can be directly  checked by observing the Moon's shadow. 

 The pointing of the array is inferred from the position of the Moon's shadow by 
high energy cosmic rays ($> 20$ TeV) which are negligibly affected by the geomagnetic field. 
This estimation can also be done by examining the deviations of the Moon's
 shadow in the north-south direction, since the effect of the geomagnetic field acts
only in the east-west direction.  It is then found to be smaller than 0.1$^\circ$ 
for both arrays.

\begin{figure}
\plotone{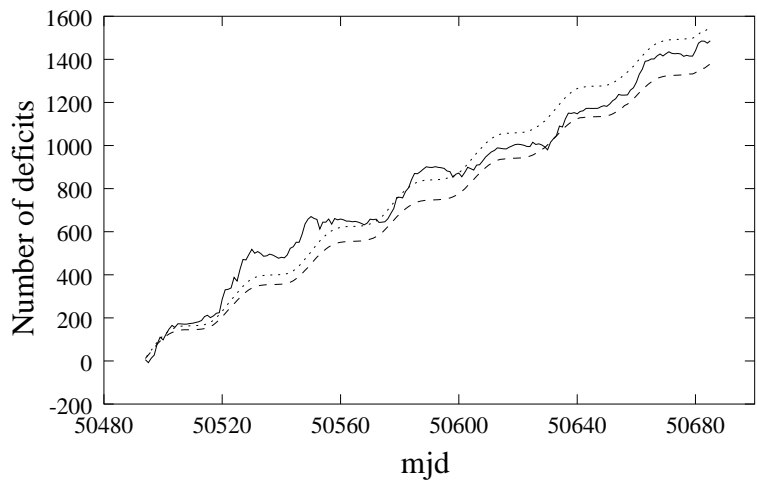}
\caption{Cumulative deficit counts of events with $\sum\rho >$ 15 coming from 
the direction of the Moon as a function of MJD (solid line).
 The radius of search window is taken to be 0.9$^\circ$ and its center is
put on the most deficit posittion of the Moon shadow. The dotted and dashed lines 
denote the expected curves when the angular resolutions are assumed to 
be 0.8$^\circ$ and 0.9$^\circ$, respectively. \label{fig2}}
\end{figure}

\placefigure{fig2}

Figure 2 shows the cumulative deficit counts of the events coming from the direction
of the Moon as a function of MJD, obtained  with the HD array. The data set used are 
between February 1997 and August 1997, just corresponding to the observation period of 
Mrk 501.  A linearly increasing of the deficit events may be  a sure  guarantee 
against the long-term stability of the array operation. 
Th meridian zenith angle of the Moon at Yangbajing changes between 12$^\circ$ and  50$^\circ$
every 27.3 days. Naturally the most efficient observations are done when the Moon 
comes in sight around the smallest zenith angle of about 12$^\circ$ every 27.3 days. 
This effect will be found as a tier-like  structure on the deficit curve, as seen in Fig. 2.

\section{RESULTS AND DISCUSSIONS}

 A circular window  was used to search for signals and then its size  was  determined 
based on the  angular resolution estimated by the experiment.
 The window size is chosen to optimize the significance of signals defined by 
$N_s/N_B^{1/2}$, where $N_S$ is the number of signals and $N_B$ the number of
background events, and to contain  more than 50~\% of the signals from a source. 
The radii of search windows used for the events with $\sum\rho >$ 15, 50 and 100
were  0.9$^\circ$, 0.8$^\circ$  and 0.5$^\circ$, respectively.
The signals were searched for by counting the number of events coming  from the 
on-source window.  The background was estimated  by averaging over events falling 
in the ten off-source  windows adjacent to the source, but without  overlapping
each other. 
The source window traverses a path in local coordinates expressed by the zenith angle 
and azimuth angle through every day. In order to reduce a strong zenith angle dependence 
of the background, 
these off-source windows were taken in the azimuth angle directions with the same zenith 
angle, except two  windows adjacent to the on-source window.

\begin{figure}
\plotone{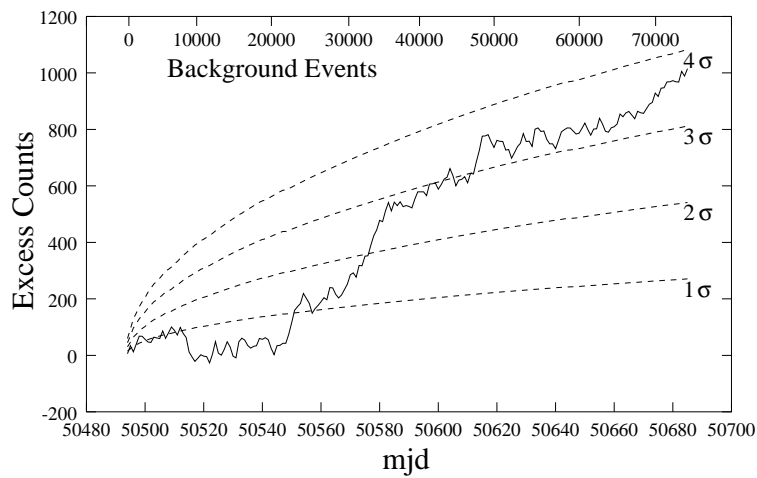}
\caption{Cumulative excess of the events with $\sum\rho >$ 15. The dotted lines
denote the excess counts at the 1, 2, 3 and 4~$\sigma$ level, respectively.
 \label{fig3}}
\end{figure}

\placefigure{fig3}

Figure 3 shows the cumulative excess counts for all events  as a function of MJD
and background, obtained with
the HD array. No excess counts were observed  until the middle of March 1997. However,
 excess events  rapidly increased  in the period from  April through June and then it
  became slightly dull. The operation of the array was stopped on August 25 of 1997 to
 calibrate the operation system.  As discussed in  \S 3, one should first note that 
the observed excess counts are by no means due to some artificial noise or unstable 
operation of the system.
The statistical significance of the excess counts  reached a 3.7~$\sigma$ level during 
this period. The excess counts very rapidly increased during the period from  April 7 
through June 16 and  the statistical significance of the excess counts was a 4.7~$\sigma$.
  These observed features are almost consistent with other observations by atmospheric Cerenkov 
telescopes (\cite{prot97}).

\begin{figure}
\plotone{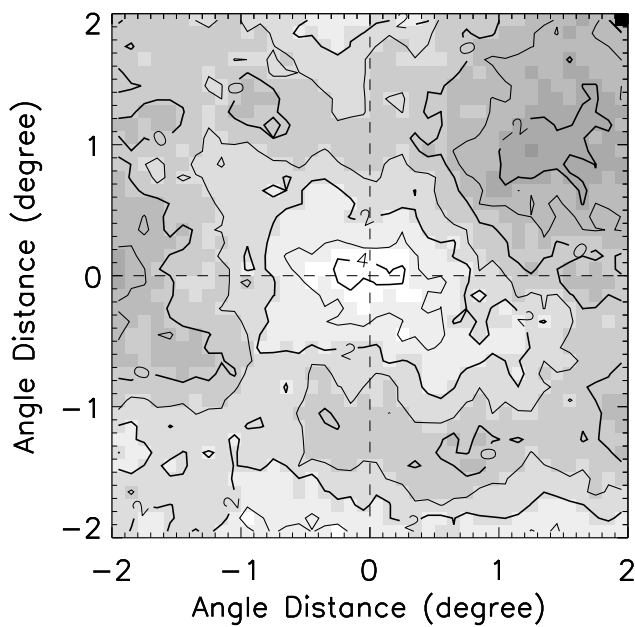}
\caption{Contour map of the weights of excess event densities around
Mrk 501, observed between April 7 and June 16 in 1997,  in the area of $4^\circ \times 4^\circ$
centered on the direction of Mrk 501.  The contour lines are drawn with a step of 
1$\sigma$.  Angle distance is measured from the direction of Mrk 501 along the right 
ascension (abscissa) and the declination (ordinate). \label{fig4}}
\end{figure}

\placefigure{fig4}

Shown in Fig. 4 is the contour map of the excess event densities around Mrk 501 for
 the events with $\sum\rho >$ 15 observed between April 7 and June 16 in 1997. 
This map was obtained using the same method as done for the Moon and Sun shadows 
(\cite{ame93}). Mrk 501 is well observed in the right direction by our air shower array. 

\begin{figure}
\plotone{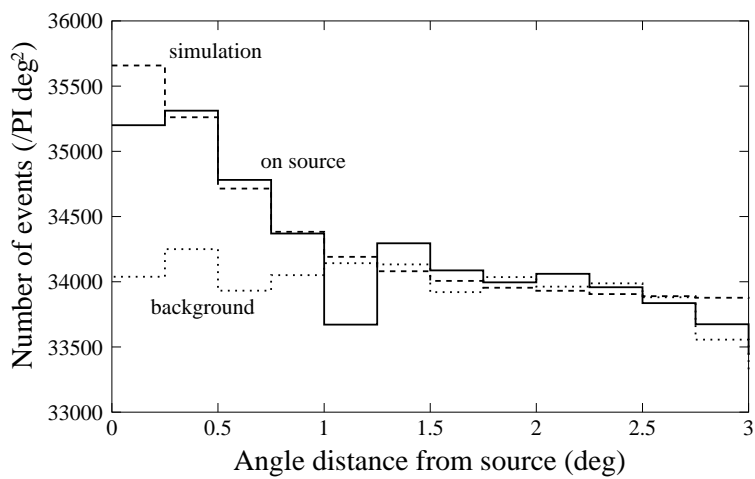}
\caption{Opening angle distribution of the events with $\sum\rho >$ 15
coming from the directions around Mrk 510. The simulation result done
for $\gamma$-ray events is shown by the dashed line. \label{fig5}}
\end{figure}

\placefigure{fig5}

 Figure 5 shows the distribution of the opening angles relative to the Mrk 501 direction
for all events with $\sum\rho >$ 15 in the HD array. The excess in the small opening angle 
region (less than 0.5$^\circ$) could be attributed  to  $\gamma$-rays from Mrk 501. 
The simulation result done for $\gamma$-ray events coming from Mrk 501 can well reproduce
 the experiment as shown in Fig.~5, when we take account of the systematic pointing errors 
estimated in \S 3.
For the observation period from 1997 February to 1997 August, the statistical 
significances of the excess events with $\sum\rho >$ 15, 30 and 50
were 3.7~$\sigma$, 2.3~$\sigma$ and 1.6~$\sigma$, respectively.

 We also searched for $\gamma$-ray emission using the entire Tibet-II array, but no excess 
was found in this period and upper limits on the excess number of the events at the 
90 \% confidence level were obtained.

We  estimated the $\gamma$-ray spectrum from Mrk 501 by a Monte Carlo simulation 
(we used a GENAS code by Kasahara \& Torii (\cite{kasa91}).), assuming 
a differential power-law spectrum with the form $E^{-\beta}$  and the
cut-off at a certain energy, $E_c$, where the cut-off means that the spectral slope  
steepens by 1.0 at $E_c$.  The value of $\beta$ was changed between 2.4 and 2.7 and 
also the effect of $E_c$ was examined  between 7 TeV  and 30 TeV. 
Primary $\gamma$-rays with energies between 0.2 TeV and 50 TeV  were thrown from 
the direction of Mrk 501.  Observation of simulated events at Yangbajing level was done
as  in our experiment, estimating the collecting area, trigger efficiency and
threshold energy for $\gamma$-rays generating the events at observation level. 
  Simulated events in respective size ($\sum\rho$) bins were then compared with those 
by the experiment.
The energy of $\gamma$-rays was defined as the energy of the maximum flux of
simulated events observed in each size bin.  These steps were repeated
until the observed results are well reproduced.  A combination of $\beta \cong 2.6$
and $E_c \sim$ 20-30 TeV can reproduce the data well.  We examined that the 
absolute flux values except the highest energy bin stay  almost unchanged for above trials, 
but  it is of course difficult to settle the spectral slope  from this experiment 
because of  very small energy range fitted here.  The  systematic errors on the flux arise
mainly from the event selection procedure, which  depends upon the array performance, 
and from the calculations of the collecting area and the air shower size 
distribution by the simulation. They are estimated to be 13 \% and 8 \%, 
respectively (\cite{ame99}).

\begin{figure}
\epsscale{.6}
\plotone{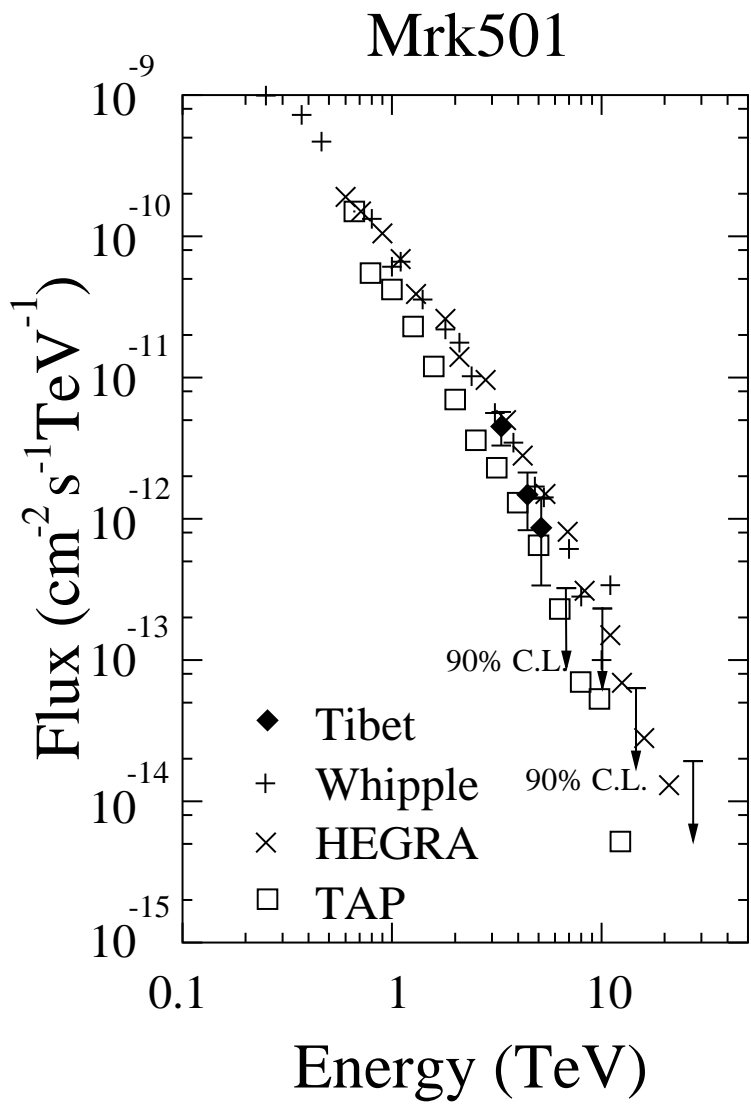}
\caption{Energy spectrum of $\gamma$-rays from Mrk 501 averaged in 
the period from 1997 February 15 to 1997 August 25. The error bars 
indicate 1~$\sigma$ ranges, excluding systematic errors.
Upper limits  at the 90~\% 
confidence level, obtained from the Tibet-II and HD arrays, are also plotted in this 
figure. Our data are compared with other results by Whipple (Samuelson et al. 1998 ),
HEGRA (Konopelko et al. 1999) and TAP (Hayashida et al. 1998). \label{fig6}}
\end{figure}

\placefigure{fig6}

\begin{figure}
\epsscale{.6}
\plotone{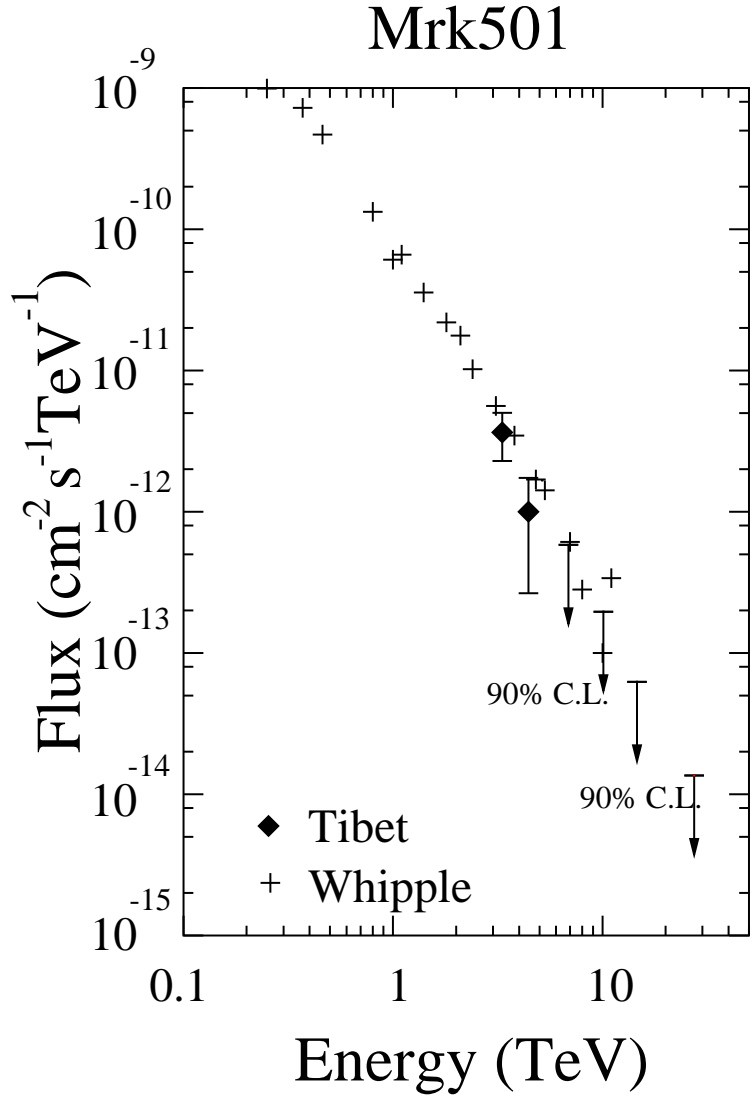}
\caption{Energy spectrum of $\gamma$-rays from Mrk 501 averaged in the period from 
1997 February 15 to 1997 June 9. Upper limits, obtained from the Tibet-II and HD arrays,
 are  at the 90~\% confidence level.   Our data are compared with the 
Whipple results. \label{fig7}}
\end{figure}

\placefigure{fig7}

 Shown in Figs. 6 and 7 are the energy spectra averaged in the period from February 
15 to August 25 in 1997 and  from February 15  to June 8 in 1997, respectively.
The latter observation time corresponds to that of the Whipple Collaboration (\cite{samu98}).
It is seen that the results reported recently by other experiments (\cite{samu98} 
; \cite{haya98} \& \cite{kon99}) are almost compatible with ours,
although these do  not cover the same observation times. It should pay attention, 
however, that our results were obtained by the continuous observation of Mrk 501 
extending  February through August in 1997, while those by Cerenkov telescopes
 were obtained for very limited  periods  of  moonless and cloudless nights.
 
 Mrk 501 and Mrk 421, nearby AGNs, are at  almost the same red-shift (0.033 and 0.031, 
respectively) and have been detected  in TeV energies (\cite{ong98}). 
In particular, Mrk 501 during the strong, long-lasting 1997 flare provided a good 
opportunity to study the energy spectrum of $\gamma$-rays from this source in detail 
(\cite{prot97}),  suggesting  a  spectral feature different with that of Mrk 421 (\cite{kren99}). 
In both sources, it is likely that a synchrotron-inverse Compton picture plays an important 
part (\cite{prot97}).   Since the attenuation mechanism of TeV $\gamma$-rays by intergalactic 
infra-red photon field  is almost the same for both sources, a  difference of spectral 
features, if any,  could be attributed to the production mechanism of $\gamma$-rays at 
the sources. Therefore, it is very important to continue  the observation  of high energy 
 $\gamma$-rays  from both sources with as small uncertainties as possible.
 
\section{SUMMARY}

 Mrk 501 suddenly came into a very active phase from March in 1997,  with several large 
flares and lasted for $\sim 1/2$ yr. The maximum $\gamma$-ray flux during this period 
reached  about 10 times as high as the Crab Nebula.
Following a  successful observation of steady emission of multi-TeV $\gamma$-rays 
from the Crab(\cite{ame99}), we further detected  multi-TeV $\gamma$-rays from Mrk 501
which was in a high flaring state between March 1997 and August 1997,  and estimated  
the absolute fluxes of $\gamma$-rays around multi-TeV region,
using the high resolution Tibet air shower array. The detection of a signal from this
source was achieved by the improvement of the array performance, which can be directly
checked by observing the Moon's shadow. Monthly observations of the Moon's shadow provide
a direct check of the angular resolution, pointing accuracy, and
also the stable operation of the array over a long period. Furthermore, the
observation of the displacement of the Moon's shadow due the effect of
the geomagnetic  field
provides an important check of the results obtained by assigning energies to all the
events. This is the first attempt to be done in the air shower experiments,
and it suggests that the Moon is a unique cosmic-ray anti-source capable of calibrating
the array  performance thoroughly. 
Hence, the results obtained by  the Tibet 
experiment using a different  technique  will be a great help  to 
understand the possible bias and errors involved  in the Cerenkov observations. 

The area of the present HD array will be extended by a factor of about five in 1999, while its
effective area will be increased by a factor of about seven by the reduction of edge effects.
 Then, the Tibet array could cover the energy range from  3~TeV to 
$\sim$100~TeV with 
 significantly better statistics and angular resolution at high energies.
 Air shower arrays are wide aperture and high duty cycle instruments, in contrast
to atmospheric Cerenkov telescopes with relatively narrow fields of view and small duty
cycle of $\sim$10 \%. These features will be indispensable for  understanding a time
variability of emission of high energy $\gamma$-rays from point sources such as AGNs
and GRBs (gamma ray bursts).
The Tibet experiment, therefore,  will have unique capabilities for the discovery of new, relatively bright sources and for a general survey of the overhead sky.

\acknowledgments

  This work is supported in part by Grants-in-Aid for Scientific
Research and also for International  Science Research from the Ministry
of Education, Science, Sports and Culture in Japan and for International
Science Research from the Committee
of the Natural Science Foundation and the Academy of Sciences in
China.

\clearpage

\end{document}